\documentclass[12pt,preprint]{aastex}

\begin{document}

%
%
\def\msun{${\rm M_{\odot}} \;$}
\def\be{\begin{equation}}
\def\ee{\end{equation}}
\def\gcc{gcm$^{-3}$}
\def\bi{\begin{itemize}}
\def\ei{\end{itemize}}
\def\Mo{$M_{\odot}$}
\def\Lo{$L_{\odot}$}
\def\s{s$^{-1}$}

\title{X-ray line tomography of AGN-induced motion in clusters of galaxies}

\author{M. Br\"uggen\altaffilmark{1}, M. Hoeft\altaffilmark{1}, M. Ruszkowski\altaffilmark{2,3}}

\altaffiltext{1}{International University Bremen, Campus Ring 1, 28759 Bremen, Germany}
\altaffiltext{2}{JILA, Campus Box 440, University of Colorado at Boulder, CO 80309-0440}
\altaffiltext{3}{Chandra fellow}

\begin{abstract}

The thermal broadening of emission lines of heavy ions is small enough
such that Doppler shifts due to bulk motions may be detected with the
next generation of X-ray observatories. This opens up the possibility
to study gas velocities in the intra-cluster medium. Here we study the
effect of bulk motions induced by a central active galactic nucleus
(AGN) on the emission lines around the FeXXV complex. We have modelled
the evolution of AGN-induced bubbles in a realistic cosmological
framework and studied the resulting FeXXV line profiles. We found that
in clusters with AGN feedback, motions induced by the inflation of
bubbles and their buoyant rise lead to distinct features in the iron
emission lines that are detectable with a spectral resolution of
$\sim$ 10 eV. These observations will help to determine the mechanical
energy that resides in the bubbles and thereby the kinetic luminosity
of the AGN.

\end{abstract}

\keywords{galaxies: active - galaxies: clusters:
cooling flows - X-rays: galaxies}

\section{Introduction}

Feedback by active galactic nuclei (AGN) is widely believed to be
relevant for the baryonic part of structure formation. Moreover, it is
likely to play a role in explaining the lower than expected mass
deposition rates in cooling flows (e.g. \citealt{fabian:03a},
\citealt{hoeft:04} and references therein). Because the interaction
between AGN and ambient medium is more easily observed in clusters of
galaxies than in galaxies, clusters serve as laboratories for the role
of AGN feedback in galaxy formation.

However, the operation of AGN feedback depends crucially on the nature
of the interaction between the AGN and the intra-cluster medium
(ICM). In recent years, high-resolution X-ray observations of cooling
flow clusters with Chandra have revealed a number of clusters that
show characteristic holes in the surface brightness. Many of these
depressions of the brightness are coincident with patches of radio
emission, e.g. in Abell 2052 (\citealt{blanton:01}), Perseus
(\citealt{boehringer:93}, \citealt{churazov:00}, \citealt{fabian:00}),
Cygnus A (\citealt{wilson:00} and \citealt{carilli:94}), Abell 2597
(\citealt{mcnamara:01}), MKW3s (\citealt{mazzotta:02}), RBS797
(\citealt{schindler:01}), HCG62 (\citealt{vrtilek:00}) and Centaurus A
(\citealt{saxton:01}). In a recent compilation Birzan et al. (2004)
lists 18 well documented clusters which show X-ray cavities with radio
emission. The consensus on these observation is that the holes are
caused by hot gas from an AGN that has displaced the ICM in these
regions.

AGN have been observed to induce sonic motions in the intra-cluster
medium which are believed to eventually dissipate to produce heat
(\citealt{fabian:03}, \citealt{ruszkowski:04}).  Here, we simulate
observations of emission lines of heavy ions that would reveal these
motions directly. Such observations would enable us to gain a better
understanding of how the kinetic energy is dissipated in the ICM. For
heavy ions such as iron, the thermal velocity is a fraction of the
speed of sound, i.e. $v_{\rm therm}\sim (m_p/m_{\rm Fe})^{1/2} c_{\rm
s} \sim 0.13c_{\rm s}$ where $c_{\rm s}$ is the speed of sound, $m_p$
the mass of the proton and $m_{\rm Fe}$ the mass of the iron
nucleus. Consequently, the thermal broadening is small enough such
that Doppler shifts of lines due to bulk motions may be detected. The
ratio between the Doppler shift of a line due to large-scale motion
with velocity $v$ and the thermal width

\begin{equation}
\Delta\nu_{\rm th} = \nu_0  \sqrt{\frac{2 k_{\rm B} T}{m_{\rm Fe} c^2}} \ , 
\end{equation}
is given by

\begin{equation}
\frac{\Delta \nu_{\rm Doppler}}{\Delta \nu_{\rm th}} \sim 3.0\, \left (\frac{v}{300 {\rm km/s}}\right ) \left (\frac{3\, {\rm keV}}{k_{\rm B}T} \right )^{1/2} \ ,
\end{equation}
where $\nu_0$ is the frequency of the emission line and $T$ the
local temperature of the ICM. All other symbols have their usual
meaning. Doppler shifts of emission lines may be detected as long as
the ICM is optically thin with respect to the line
considered. Resonant scattering can increase the optical
depth. However, \citet{churazov:04} have performed XMM observations of
the Fe K-lines near 6.7 keV in the Perseus cluster and concluded that
resonant scattering may be suppressed by strong turbulence. Thus the
velocities in the cluster center keep the medium optically thin, and
consequently, the shapes and positions of the lines may reveal bulk
motions in the ICM. The study of iron line profiles has previously
been suggested as indicators for turbulence in the ICM by
\citet{inogamov:03} and \citet{sunyaev:03}.

Upcoming generations of X-ray observatories such as ASTRO-E2 will have
an energy resolution of $\sim 5$ eV at full width at half-maximum in
the photon range from 0.5 to 10 keV
\footnote{http://www.isas.ac.jp/e/enterp/missions/astro-e2/}. This
range includes the main lines from Fe XXV at energies around 6.7 keV
which we are going to use as an example to study the effect of
macroscopic motions.  For the case of Perseus, ASTRO-E2 will be able
to resolve the Fe-K line with a precision of better than 100 km/s. The
XEUS mission\footnote{http://www.rssd.esa.int/index.php?project=XEUS\&page=index}
is going to achieve an energy resolution of $\sim 1$ eV at spatial
resolutions of a few arcseconds. This provides the opportunity to map
the detailed velocity structure at least in the nearby clusters.

We have simulated the dynamics of buoyant bubbles in a realistic
cluster using the AMR (adaptive mesh refinement) code FLASH. In this
paper we study the exemplary spectra of helium-like iron lines along a
variety of lines of sight through the cluster to search for signatures
of motions induced by bubbles.

\section{Simulation}

Numerical simulations of hot, underdense bubbles in clusters of
galaxies have been performed by a number of authors
(e.g. \citealt{churazov:01}, \citealt{quilis:01},
\citealt{bruggen:02}, \citealt{bruggen:02a}, \citealt{reynolds:02},
\citealt{ruszkowski:04}, \citealt{ruszkowski:04b},
\citealt{dallavecchia:04}).  Common to these simulations is that they
use a spherically symmetric, analytical profile for the ICM. However,
for the purpose of studying the spectral features of AGN we were
concerned that such a sterile environment without the typical motions
expected in a cluster might result in unrealistically clean
spectra. Therefore, we produced a three-dimensional hydrodynamical
simulation of bubbles in a cluster that has been extracted from a
cosmological simulation (similar to the one reported in
\citet{bruggen:05}). This simulation was computed with the
Smooth-Particle Hydrodynamics (SPH) code GADGET in standard
$\Lambda$CDM cosmology ($\Omega_{\Lambda}= 0.7$, $\Omega_{\rm m}=
0.3$, $h= 0.7$) and is a re-run of the S2 cluster in
\citet{springel:01b}. At redshift $z=0$ this cluster has a mass of
$7\cdot 10^{14}$ \Mo, a central density of $n_{\rm e}\sim 0.1$
cm$^{-3}$ and a central temperature of 6 keV. This cluster was
chosen because its total mass and central temperature were closest to
the Perseus cluster ($T\sim 6.3$ keV and mass $\sim 1\cdot 10^{15}$
\msun). The SPH simulation of the cluster includes radiative cooling
and star formation. The output of the SPH simulation serves as initial
model for our AMR simulation.

FLASH is a modular block-structured AMR code, parallelised using the
Message Passing Interface (MPI) library. It solves the Riemann problem
on a Cartesian grid using the Piecewise-Parabolic Method (PPM) and, in
addition, includes particles that represent the collisionless dark
matter and stars. In our simulation 714346 particles were
used. The particles are advanced using a cosmological
variable-timestep leapfrog-method.  Gravity is computed by solving
Poisson's equation with a multigrid method using isolated boundary
conditions. For the relatively short physical time of the bubble
simulation, radiative cooling and star formation are neglected, even
though they were included in the cosmological SPH simulation with
which the cluster was produced. The computational domain of our AMR
simulation is a cubic box of side $L=2 h^{-1}$ Mpc. For our grid, we
chose a block size of $16^3$ zones and used outflow boundary
conditions. The minimal level of refinement was set to 3 which means
that the minimal grid contains $16 \cdot 2^{(3-1)} = 64^3$ zones.  The
maximum level of refinement was 7, which corresponds to an effective
grid size of $16 \cdot 2^{(7-1)} = 1024$ zones or an effective
resolution of $1.96 h^{-1}$ kpc. The code was run on 64 processors of
the IBM p690 at the John-von-Neumann Institute for Computing in
J\"ulich, Germany. A single run took approximately 5000 CPU
hours.

\begin{figure}[htp]
\begin{center}
\includegraphics[width=1.0\textwidth]{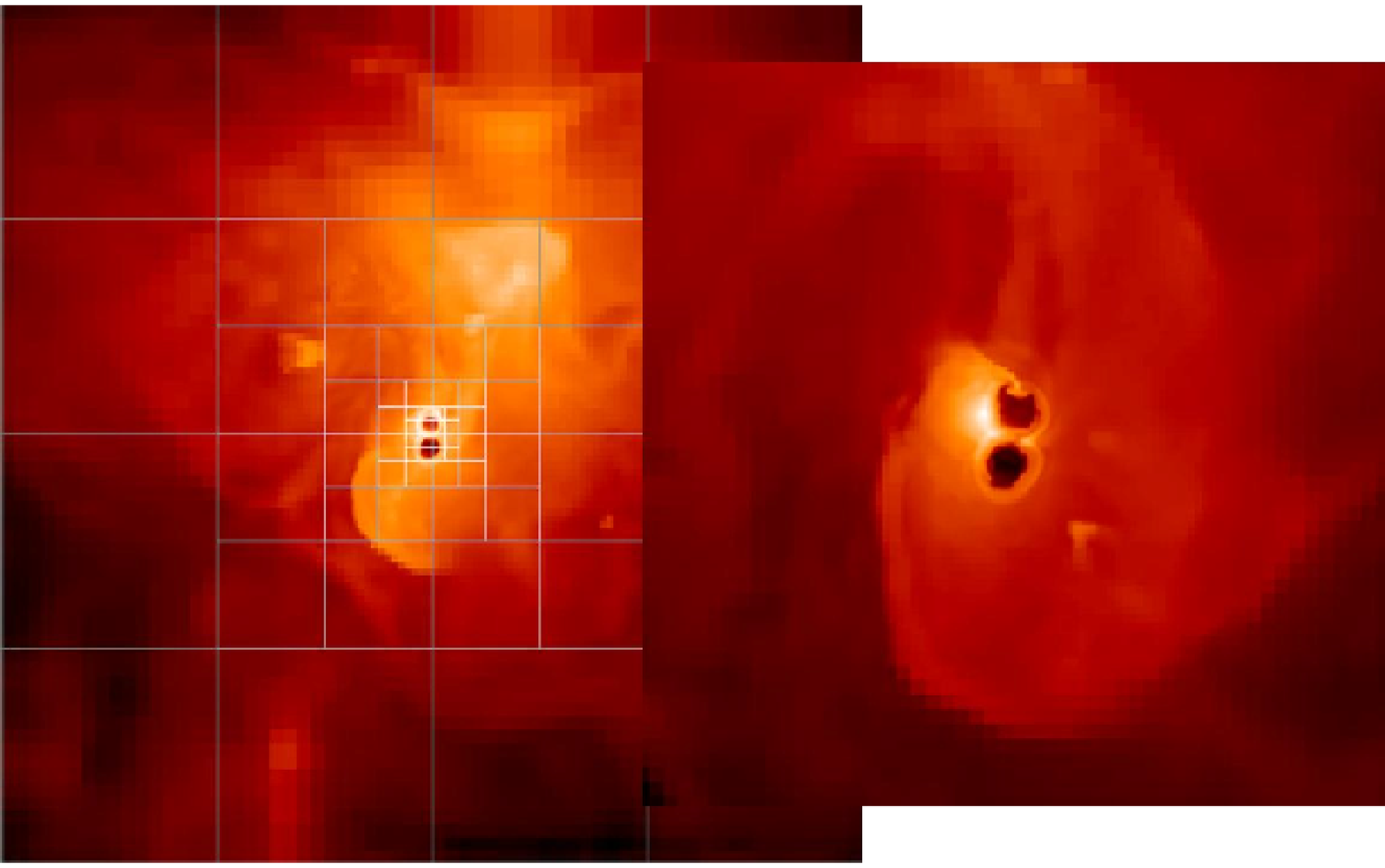}
\end{center}
\caption{Contour plot of the gas density in a slice through the center
of the cluster at a time of $t=70$ Myrs after the start of AGN
activity. The left slice represents 2 $h^{-1}$ Mpc a side, the inlet
shows an enlargement of the inner 400 $h^{-1}$ kpc.}
\label{fig1}
\end{figure}

Fig. \ref{fig1} shows the block boundaries in a slice displaying the
gas density at a time of 70 Myr after the start of AGN activity. At
this stage the bubble has not yet achieved pressure equilibrium with
its surroundings and is still expanding nearly spherically into the
ambient medium. The bubbles are nearly spherical and do not look
dissimilar from the bubbles observed in the Perseus cluster and other
clusters. As a result of its rapid inflation, it has produced a weak
shock wave that has started to travel outwards from the bubbles. This
shock wave is clearly visible in the density plot
(Fig. \ref{fig1}). The same figure also shows that the cluster is
quite dynamic and shows significant substructure such as clumps and
shock fronts.

The AGN is assumed to sit in the center of the cluster. Energy is
injected into two spherical regions that lie at a radial distance of
30 kpc on either side of the center of the cluster. The energy is
injected by increasing the specific internal energy within the
injection regions at a constant rate. For a period of $3.5\cdot 10^7$
years a total energy of $10^{60}$ erg per bubble is injected into two
spherical regions of radius 13 kpc. These parameters were chosen
because they produced bubbles that are very similar to those observed
in the Perseus cluster.

\section{Results and Discussion}

%

Here we investigate the spectral signatures of AGN-induced motions at
the example of the complex near the FeXXV K$_\alpha$ line. The spectra
are computed using the {\sc ATOMDB} database
(http://cxc.harvard.edu/atomdb/). For each temperature {\sc ATOMDB}
determines the ionization fractions and the emissivity. The emissivity
from each computational cell is obtained (i) by convolving the line
spectrum with a Gaussian of width $\Delta \nu_{\rm th}(T)$ to model
the thermal broadening and (ii) by Doppler-shifting the line by the
corresponding velocity component $ \Delta\nu_v = \nu_0 \, v / c $.

The spectral surface brightness is determined by integrating along the
line of sight
\begin{equation}
        I(\nu)
        =
        \int_{\rm l.o.s.}
        {\rm d}x \  \:
        n_{\rm e}(x)
        n_{\rm Fe}(x) \:
        \int
        {\rm d}\nu' \:
        \frac{1}{\sqrt{\pi} \Delta\nu_{\rm th}(x)}
        \exp
        \left\{
                - \frac{ (\nu - \nu'- \Delta\nu_z - \Delta\nu_v )^2 }
                       { \Delta\nu_{\rm th}^2(x) }
        \right\}        \:
        j( T(x), \nu' - \Delta\nu_z - \Delta\nu_v )
        ,
        \label{eq-los}
\end{equation}
where $n_{\rm e}$ is the electron number density, $n_{\rm Fe}$ the
number density of iron atoms and $j$ the emissivity.  Here, we neglect
the effect of the cosmological redshift since the redshift difference
of light emitted from the front and the back of the cluster is small
compared to bulk motion redshift. The bulk motions caused by
cluster mergers or -- as considered here -- a central AGN may cascade
down to sub-grid turbulent motions.  This small-scale turbulence
produces additional line broadening as described in
\citet{inogamov:03} and \citet{sunyaev:03}.  Here we ignore Doppler
broadening below the grid scale since it is small compared to the
Doppler shifts produced by the macroscopic motions (see
\citet{inogamov:03}). Thus, the line width of gas within a single cell
is given solely by the thermal width: $\Delta\nu_{\rm th}$, see
Eq.~(1).

\citet{finoguenov:00} investigated the heavy element distribution in
clusters and found that iron is generally more abundant in the central
regions. Since our single-fluid simulation does not trace any chemical
enrichment history, we need to model the iron abundance. Several
clusters that resemble our model cluster, show a decrease in their
iron abundance beyond $\sim 100\:{\rm kpc}$. Only clusters that have
suffered a recent merger have a more uniform iron abundance. Cluster
A3112 has a temperature of $T_e = 5.3\:{\rm keV}$, hosts a powerful
radio source in its center and is believed to have a strong cooling
flow. Hence, we chose it as a model for our iron abundance profile
which we approximate by
\begin{equation}
        f_{\rm Fe}
        =
        \frac{ r_{\rm Fe}}
        { r_{\rm Fe} + r }      
        ,
\end{equation}
with $r_{\rm Fe} = 300\:{\rm kpc}$, which is the radius at which the
iron abundance in A3112 drops to one half. This is a conservative
model because in the Perseus cluster the metals are even more
centrally concentrated \citep{schmidt:02} which would lead to an
increased sensitivity of Fe-line Doppler shifts to motions in the
cluster center.

\begin{figure}[htp]
\begin{center}
\includegraphics[width=1.0\textwidth]{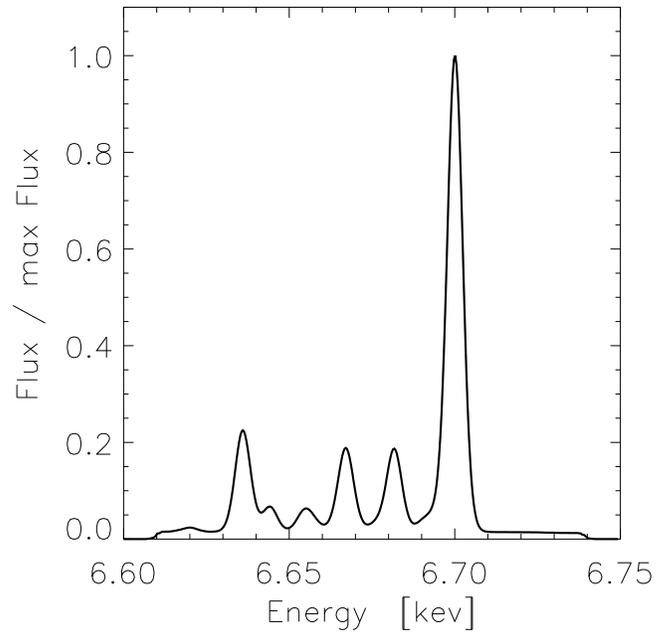}
\end{center}
\caption{A purely thermally broadened spectrum along a
line-of-sight. The most prominent line is the FeXXV K$_\alpha$
transition at 6.7~keV.}
\label{fig2}
\end{figure}

For reference Fig. \ref{fig2} shows a spectrum in the vicinity of the
FeXXV K$_\alpha$ transition. This spectrum is merely thermally
broadened and ignores all hydrodynamic Doppler shifts. It is computed
along a line of sight through the center of the cluster before the
onset of AGN activity.

\begin{figure}[htp]
\begin{center}
\includegraphics[width=0.8\textwidth]{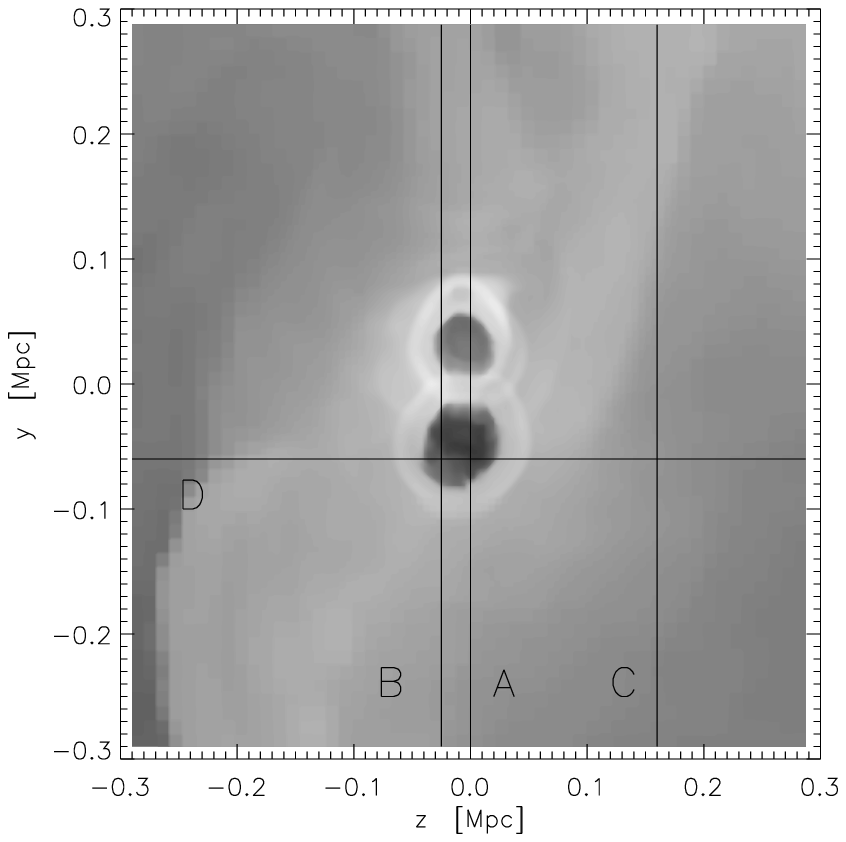}
\includegraphics[width=0.6\textwidth]{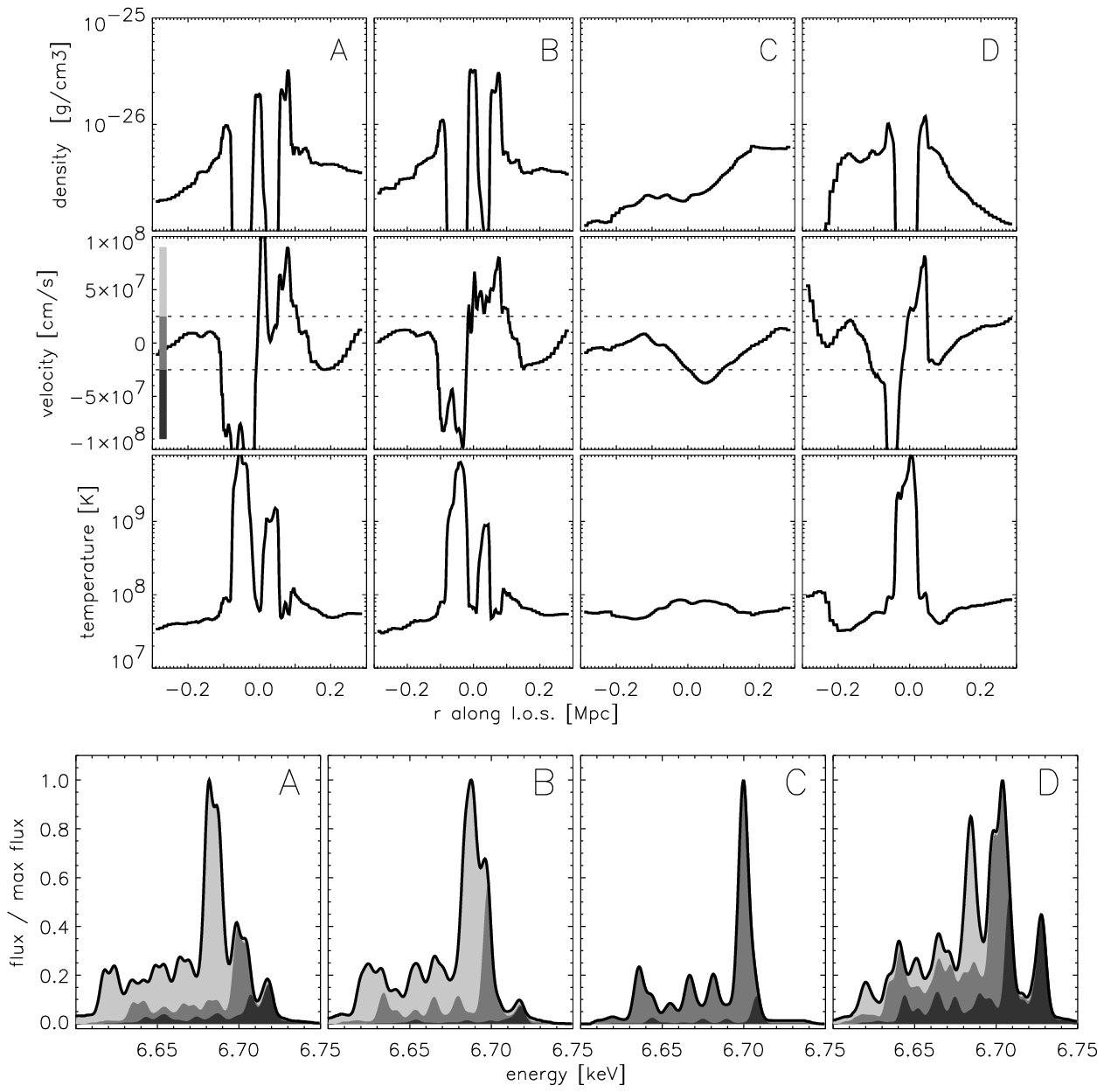}
\end{center}
\caption{Spectra computed along different lines of sight through the
cluster. The upper panel shows the density in a slice through the
cluster center at a time $70~{\rm Myr}$ after the start of AGN
activity. The four lines of sight for which spectra are shown in the
lower panel are overplotted (labelled A-D). The density, velocity, and
temperatures along the lines of sight are depicted in the upper rows
in the lower panel. The bottom row shows the corresponding
spectra. Here the light gray area shows the contribution to the total
spectrum coming from gas whose velocity lies above the upper dashed
line in the velocity profile, the dark gray area corresponds to gas below
the lower dashed line in the velocity profile and the medium gray area
corresponds to gas in the middle part. Finally, the bold line shows
the total spectrum.}
\label{fig3}
\end{figure}

Now we wish to compare this spectrum to one that includes hydrodynamic
motions. Fig. \ref{fig3} shows a density contour plot with some
exemplary lines of sight (labelled A-D) which intersect the cluster at
various impact parameters and from various directions. Below are the
corresponding profiles (density, l.o.s. velocity and temperature) and
spectra. For those l.o.s. that cross a bubble the density profiles
clearly shows the underdense regions. L.o.s. A and B intersect both
bubbles, while line D intersects only one bubble in a direction
perpendicular to the jet axis; finally line C is significantly off-set
from the center and removed from all AGN activity. The underdense
bubbles are surrounded by shells of compressed material with
temperatures of about $10^8\:{\rm K}$, which gives rise to significant
FeXXV emission. At the same time, these shells show the highest
velocities which are in excess of $1000\:{\rm km\,s^{-1}}$. As a
result the most prominent emission line shows a clear triplet
structure, as for example in line A: On top of the unshifted
emission coming from regions removed from the bubble regions, there
are both a red-shifted and a blue-shifted line. The resulting
separation of the two shifted lines is approximately $35\:{\rm
eV}$. In order to identify which parts of the spectrum are produced by
which regions, we have split up the spectrum according to which
velocity the emitting gas has. In Fig. \ref{fig3} the light gray area
shows the contribution to the total spectrum coming from gas whose
velocity lies above the upper dashed line in the velocity profile, the
dark gray area corresponds to gas below the lower dashed line in the
velocity profile and the medium gray corresponds to gas in the middle
part. Finally, the bold line shows the total spectrum. Thus, we can
see that the blue-shifted line comes from the lower rim of the gas
which has large negative velocities (indicating motion towards the
observer) and the red-shifted peak comes from the rim which has large
positive velocities. The resulting effective broadening of the line is
greater than 35 eV and will thus be detectable with the next
generation of X-ray telescopes.

The l.o.s. labelled B with an impact parameter of $20$ kpc also
crosses both bubble rims. Again it yields a multi-line structure in
the spectrum leading to a broadening of about $20\:{\rm eV}$. Clearly,
the velocity shifts are discernible over the entire scale of the
bubbles. For comparison, l.o.s. C which is off-set from the center and
unaffected by the AGN shows very little broadening and no
splitting. The spectrum differs very conspicuously from those of lines
A and B.

The same feature is observed when the line of sight is
perpendicular to the axis of the AGN as in line D. Whenever the
l.o.s. crosses the hot, fast-moving material in the bubble rim, the
main line becomes strongly broadened or split. The broadening of the
most prominent line is even greater than for lines A and B, amounting to
to $\sim 40\:{\rm eV}$. Furthermore, the amount of compressed material in the
rim is smaller, hence the shifted features have less emission. Since
the bubble is at a fairly early phase in its evolution, the bubble
rims are expanding which leads to Doppler shifts and a marked feature
in the resulting spectrum.

\begin{figure}[htp]
\begin{center}
\includegraphics[width=0.8\textwidth]{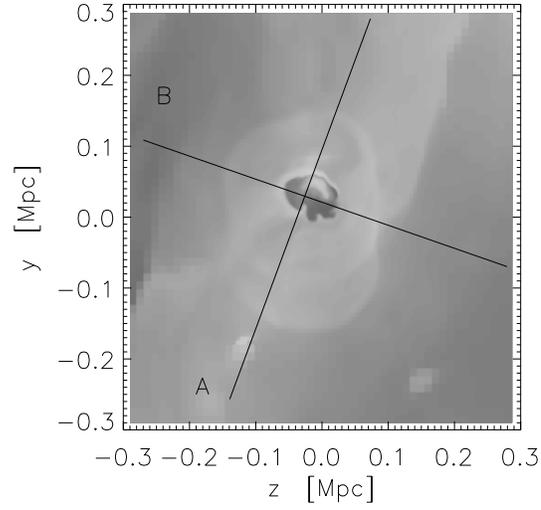}
\includegraphics[width=0.6\textwidth]{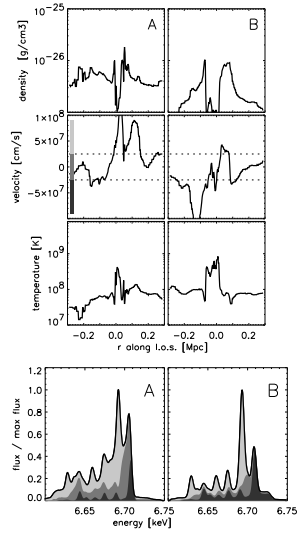}
\end{center}
\caption{
Similar to Fig.~\ref{fig3} but now for a snapshot $140~{\rm Myr}$ after start of the AGN activity. 
}
\label{fig4}
\end{figure}

\begin{figure}[htp]
\begin{center}
\includegraphics[width=0.8\textwidth]{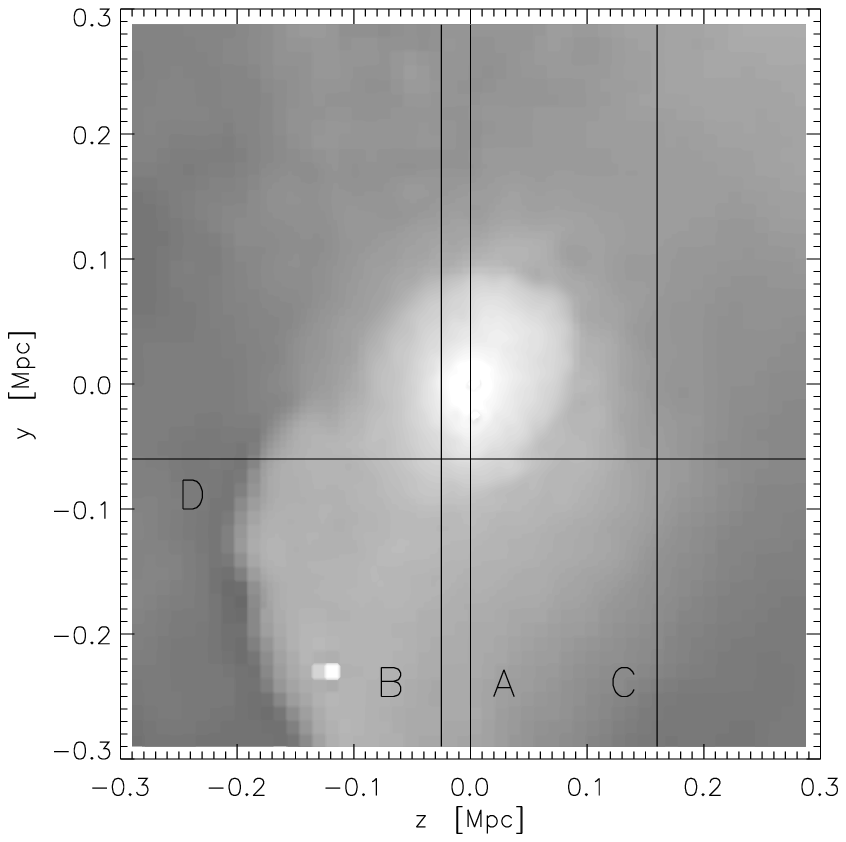}
\includegraphics[width=0.6\textwidth]{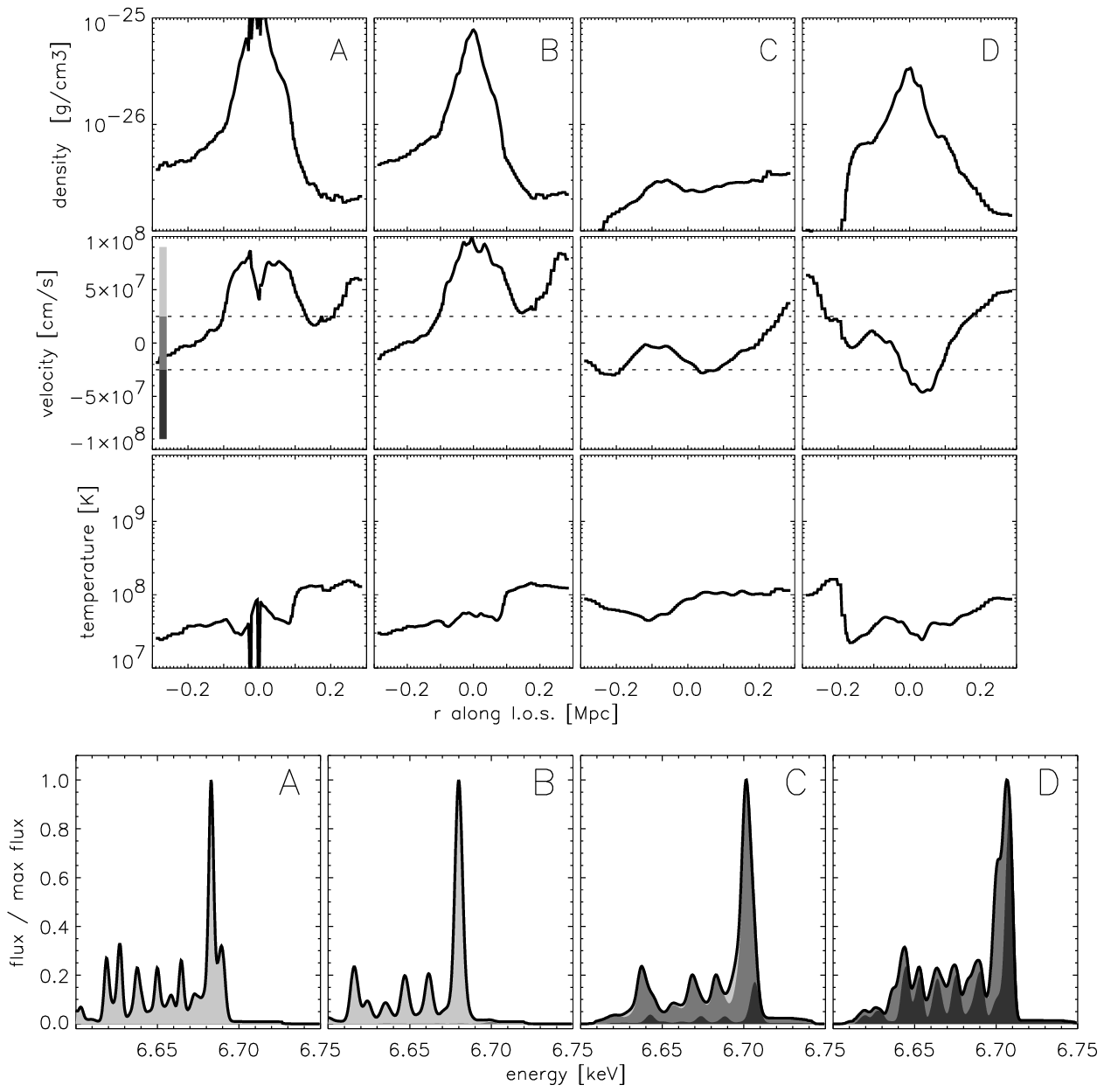}
\end{center}
\caption{
Similar to Fig.~\ref{fig3} but now for a time directly before the start of the AGN activity. 
}
\label{fig5}
\end{figure}

In Fig. \ref{fig4} we show the bubble at a later stage ($t=140$
Myrs). In the cut through the center of the cluster only one bubble is
visible now as the second one has already been advected out of the
plane, i.e. ambient motions have moved it in a direction perpendicular
to the cut through the computational domain shown in
Fig. \ref{fig4}. We show two characteristic l.o.s., A and B. Again,
both spectra show a splitting of the most prominent line. As in the
previous snapshot, these features are produced by the fast material
that is pushed out as the bubble rises and expands. Fig. \ref{fig4}
shows that virtually all l.o.s. through the bubble show strong traces
of the bubble in their spectra and that these signatures do not solely
occur in the early phase of their evolution. For comparison we
have computed spectra along identical lines-of-sight in the same
cluster but without bubbles (see Fig. \ref{fig5}). It is apparent from
comparison with Fig. \ref{fig3} that, in the absence of bubbles, the lines
are very much less affected and, consequently, that the bubble-induced
motion leave a clear imprint on the spectra.

To summarise one can cite two conditions for observing the
bubble-induced velocities via X-ray lines from heavy ions: First, the
gas needs to show sufficiently large velocities relative to the
ambient ICM. In the study presented here the inflation of the bubbles
leads to velocities of about 500 to $1000\:{\rm km \,s^{-1}}$ which is
a significant fraction of the local sound speed. Second, the
accelerated material needs to emit a sufficient amount of photons in
the chosen lines. Thus, the early phase of bubble motion is
advantageous (for FeXXV), since the adiabatically compressed
material ahead of the bubble has not had enough time to cool. As a
consequence of the compression the gas is dense and hot which, in
turn, leads to a high emissivity and a strong emission signal.
However, the effects on the spectrum are not confined to the very
early phases. They should persist for at least 100 Myrs after the
onset of AGN activity. Once the material around the rims has cooled
down significantly, we expect the effects on the spectrum to wane.

\begin{figure}[htp]
\begin{center}
\includegraphics[width=0.7\textwidth]{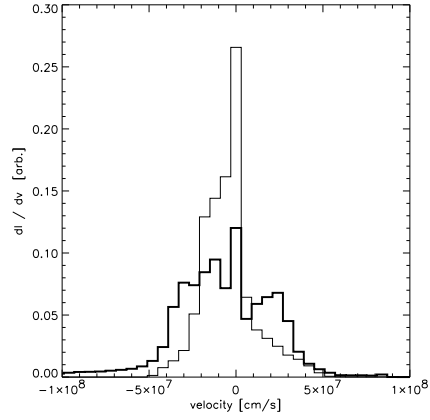}
\end{center}
\caption{Emission-weighted velocity distribution in a volume of size
60 kpc x 60 kpc x 600 kpc, parallel to the line-of-sight and centred
on a bubble. The velocities are weighted with the emissivity in the
FeXXV K$_\alpha$ line. The bold line refers to the cluster with
bubbles and the thin line to the run without bubbles. Units on the
ordinate axis are arbitrary.}
\label{fig6}
\end{figure}

In Fig. \ref{fig6} we have plotted the velocity distribution,
weighted with the emissivity in the FeXXV K$_\alpha$ line, in a volume
of size 60 kpc x 60 kpc x 600 kpc, parallel to the line-of-sight and
centred on a bubble. The bold line refers to the cluster with bubbles
and the thin line to the run without bubbles. This histogram reveals
that the presence of the AGN leads to a broadening of the velocity
distribution. This difference in the emission-weighted velocity
distribution leads to discernible differences in the spectrum, even if
averaged over a large field of view.

Besides inspecting spectra along individual lines of sight one can
attempt to quantify the effects caused by bubble motions by reducing
the line shifts to a single parameter. To demonstrate this we have
produced a map of the number of photons in the band [6.71-6.73] keV
normalised by the number of photons in a reference band between
[6.68-6.71] keV. The band between 6.71-6.73 keV was chosen because in
the absence of motions one would expect very little emission in this
band because it lies blueward of the FeXXV complex and is not
contaminated by any other lines. Thus, emission in this band is a
clear indication of blue-shifted lines from the Fe complex.

\begin{figure}[htp]
\begin{center}
\includegraphics[width=0.7\textwidth]{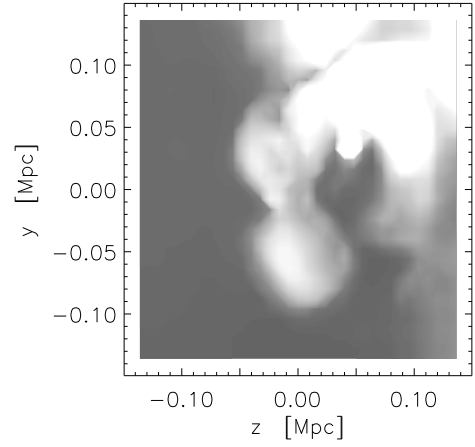}
\end{center}
\caption{The ratio between emission in two adjacent bands within the
iron complex, i.e. $I_{6.685-6.71}/I_{6.71-6.725}$. The darker
colour corresponds to a lower ratio.}
\label{fig7}
\end{figure}

Fig. \ref{fig7} shows such a map which reveals a dumbbell-shaped
imprint from the blue-shift produced by the gas pushed out radially by
the bubbles. On the upper right-hand corner one can see additional
bright features that are related to substructure near the center of
the cluster. Clearly this simple procedure is susceptible to any
effect that alters the emission ratio between the two bands.  For
instance the large-scale streaming motion of the gas can be clearly
seen from the overall gradient in Fig.~\ref{fig7}.  Therefore, it is
particularly encouraging that the bubbles leave such a distinct
imprint on the map.

The bubbles depicted in Fig.~\ref{fig7} resemble those observed in the
Perseus cluster (\citealt{fabian:03}). The bubbles in our simulation
start with a larger separation and they are already larger in size
compared to those in Perseus.  However, when our line of sight is
perpendicular to the jet axis, as it is in the case of Perseus, the
early stages of the bubble evolution should be detectable in a map
such as Fig.~\ref{fig7} since the bubbles are still
expanding. Therefore, we would expect a strong blue excess for the
innermost region within about $\sim 25$ kpc from the center.  The
temperature map of Perseus (\citealt{fabian:03}) shows that the
material around the bubbles is to some extent colder than the ICM in
average. Fig.~\ref{fig4} l.o.s. B resembles such a situation, the
bubbles are surrounded by cold material, still a significantly shifted
$K_\alpha$ line is discernible.

Since the bubble configuration presented here is similar to that in
Perseus, it may shed some light on the gas velocities typically
present in clusters with active bubble formation. \cite{churazov:04}
argued from the absence of resonant scattering that motions with
velocities half of the speed of sound are present in the cluster core.
The l.o.s. A in Fig.~\ref{fig3} indicates that the gas between the
bubbles is also highly accelerated. The same is true for gas close to
a bubble as can be seen from l.o.s B. The study of the velocity
structure around X-ray cavities would help to determine the mechanical
energy that resides in the bubbles and thereby the kinetic luminosity
of the AGN. This, again, is important for establishing how much energy
is available for dissipation that can offset cooling in the core of
the cluster.

Summary:

\begin{itemize}

\item AGN-induced motions of the ICM can significantly alter the line
shapes of X-ray lines of heavy ions. For a typical bubble in a cluster
such as Perseus these line shifts are greater than 20 eV for all lines
of sight that intersect the bubble and should therefore be observable
with the next generation of X-ray spectrographs.

\item Here we studied the line shift at the example of the FeXXV lines
around 6.7 keV. The line shifts are caused primarily by the hot
compressed rims of the expanding bubble. Other regions of the cluster
contribute relatively little to the distortions of the line. The line
shifts are therefore relatively insensitive to other motions along the
line of sight in the outer regions of the cluster. Observations of
such line shifts should allow a better determination of the energy
that resides in the bubbles and the motions they induce in the ICM.

\end{itemize}

\noindent {\sc acknowledgement }

We thank Volker Springel for providing us with some of his simulations
and Hans Boehringer for helpful comments on the paper.  MB gratefully
acknowledges support by DFG grant BR 2026/2 and the supercomputing
grant NIC 1658. MR acknowledges the support from NSF grant AST-0307502
and NASA through {\it Chandra} Fellowship award number PF3-40029
issued by the Chandra X-ray Observatory Center, which is operated by
the Smithsonian Astrophysical Observatory for and on behalf of NASA
under contract NAS8-39073.  The software used in this work was in part
developed by the DOE-supported ASCI/Alliance Center for Astrophysical
Thermonuclear Flashes at the University of Chicago. Finally, the
referee is thanked for many helpful comments.

\newpage

\bibliography{radio}
\bibliographystyle{apj}

\clearpage

\clearpage

\clearpage

\clearpage

\clearpage

\end{document}